\documentclass[conference]{IEEEtran}
\IEEEoverridecommandlockouts
\usepackage{cite}
\usepackage{amsmath,amssymb,amsfonts}
\usepackage{algorithmic}
\usepackage{graphicx}
\usepackage{textcomp}
\usepackage{xcolor}
\usepackage[letterpaper,top=0.75in,bottom=1.05in,left=1.03in,right=1.03in]{geometry}
\usepackage{url}
\usepackage{balance}
\def\BibTeX{{\rm B\kern-.05em{\sc i\kern-.025em b}\kern-.08em
    T\kern-.1667em\lower.7ex\hbox{E}\kern-.125emX}}
\IEEEoverridecommandlockouts  
\begin{document}

\title{SDAP-based QoS Flow Multiplexing Support in Simu5G for 5G NR Simulation\\
\thanks{This publication has emanated from research conducted with the financial
support of Science Foundation Ireland under Grant number 13/RC/2077 P2.
For the purpose of Open Access, the author has applied a CC-BY public
copyright licence to any Author Accepted Manuscript version arising from
this submission}
}

\author{\IEEEauthorblockN{Mohamed Seliem, Utz Roedig, Cormac Sreenan and Dirk Pesch}
\IEEEauthorblockA{\textit{School of Computer Science and Information Technology} \\
\textit{National Universities of Ireland, University College Cork}, Cork, Ireland \\
MSeliem@ucc.ie, u.roedig@ucc.ie, Cormac.Sreenan@ucc.ie, Dirk.pesch@ucc.ie}
}

\maketitle

\begin{abstract}
The Service Data Adaptation Protocol (SDAP) plays a central role in 5G New Radio (NR) by enabling QoS Flow multiplexing over shared Data Radio Bearers (DRBs). However, most 5G simulation frameworks, including Simu5G, lack SDAP support, limiting their ability to model realistic QoS behavior. This paper presents a modular, standards-compliant SDAP extension for Simu5G. Our implementation introduces QFI-based flow tagging, SDAP header insertion/removal, and configurable logical DRB mapping—fully integrated into Simu5G while preserving compatibility with LTE-based components. The proposed design supports multi-QFI simulation scenarios and enables researchers to model differentiated QoS flows and flow-aware scheduling policies. Validation results confirm correct SDAP behavior and pave the way for advanced 5G simulations involving per-flow isolation, latency-sensitive traffic, and industrial QoS profiles.
\end{abstract}

\section{introduction}
The deployment of 5G New Radio (NR) introduces a fully rearchitected radio access network designed to deliver significantly enhanced Quality of Service (QoS) flexibility compared to previous cellular generations. A key enabler of this flexibility is the Service Data Adaptation Protocol (SDAP), positioned between the IP layer and the Packet Data Convergence Protocol (PDCP) sublayer within the user plane protocol stack. SDAP supports the multiplexing of multiple QoS flows over Data Radio Bearers (DRBs), utilizing fine-grained QoS Flow Identifiers (QFIs), and facilitates per-flow differentiation while maintaining interoperability with lower protocol layers.

In simulation-based 5G research, Simu5G has emerged as a widely adopted open-source framework for evaluating 5G networks. Built atop the OMNeT++ discrete-event simulation platform, Simu5G offers comprehensive support for numerous 5G components, including the NR PHY, MAC, Radio Link Control (RLC), and PDCP layers. However, it lacks native support for the SDAP layer. Consequently, existing Simu5G simulations rely on simplified bearer models inherited from LTE, which omit SDAP header processing and QFI-based multiplexing. The lack of SDAP functionality in Simu5G limits the ability of researchers to study key 5G features such as:  QoS Flow multiplexing,  Differentiation of traffic based on QFIs, Control-plane driven DRB management, and QoS-aware scheduling and resource allocation.

In this paper, we introduce a novel extension to Simu5G that integrates full SDAP header processing and QFI-aware flow handling into the user plane protocol stack. Our design incorporates:
\begin{itemize}
    \item Standards-compliant insertion and removal of SDAP headers based on 3GPP TS 38.331 and TS 38.413;
    \item Flow-based QFI tagging using INET's packet tagging infrastructure;
    \item Configurable QFI-to-DRB mapping to support per-flow logical bearer associations;
    \item Full backward compatibility with the existing Simu5G architecture, with no required modifications to the PDCP, RLC, or MAC layers.
\end{itemize}

This implementation enables multiple applications within a simulated User Equipment (UE) to transmit using distinct QFIs while maintaining a protocol behavior that closely mirrors the 5G NR specifications. Furthermore, the modular nature of the design provides a foundation for future enhancements, including physical DRB multiplexing, dynamic control-plane procedures, and support for advanced SDAP features such as the Reflective QoS Indicator (RQI).

The remainder of this paper is structured as follows: Section \ref{section1} provides background on the SDAP protocol. Section \ref{section2} reviews Simu5G’s architecture and outlines current limitations. Section \ref{section3} presents the design and integration of our SDAP extension. Section \ref{section4} details the implementation. Section \ref{eval} reports on evaluation results. Sections \ref{section5} and \ref{section6} conclude the paper and outline directions for future work.

\pagebreak
\section{Background: SDAP in 5G NR} \label{section1}
The Service Data Adaptation Protocol (SDAP) is a newly introduced sublayer in the 5G NR user plane protocol stack, sitting between the IP layer and the PDCP layer, as specified by 3GPP in TS 38.331 \cite{b1} and TS 38.413 \cite{b2}. The primary function of SDAP is to support multiplexing of multiple QoS flows into DRBs, thereby allowing 5G NR to achieve per-flow QoS management across highly diverse traffic types\cite{b11}.

In 5G NR, each QoS flow is identified by a QFI, which is assigned by the control-plane. Multiple QFIs can be mapped to a single DRB, enabling efficient multiplexing and resource sharing. Conversely, some use cases may require certain QFIs to be assigned to dedicated DRBs for strict isolation. The SDAP layer is responsible for handling this multiplexing logic, as well as inserting and removing SDAP headers containing the QFI and additional control fields for each packet.

The SDAP header structure is relatively lightweight, typically consisting of the following fields \cite{b1}, \cite{b2}:
\begin{itemize}
    \item QoS Flow Identifier (QFI): 6-bit field identifying the QoS Flow;
    \item D/C (Data/Control) bit: indicates whether the PDU contains data or control information;
    \item Reflective QoS Indicator (RQI): enables reflective QoS mechanisms when set.
\end{itemize}

This mechanism allows 5G NR to support various QoS classes defined by 3GPP TS 23.501 \cite{b3}, including ultra-reliable low-latency communication (URLLC), enhanced mobile broadband (eMBB), and massive machine-type communication (mMTC), each with specific latency, reliability, and throughput requirements.

While 5G NR’s lower layers such as PDCP, RLC, MAC, and PHY handle segmentation, retransmissions, scheduling, and radio access control, SDAP introduces a crucial flow-level adaptation layer that did not exist in LTE networks. Notably, LTE relied on bearer-based mapping alone, without QFI or SDAP functionality.

The inclusion of SDAP in 5G is therefore a major step toward enabling dynamic, fine-grained QoS control across heterogeneous traffic profiles, which is especially critical in industrial, automotive, and mission-critical communication scenarios.

\section{Simu5G Architecture and Limitations} \label{section2}
Simu5G is an open-source, end-to-end 5G system-level network simulator developed as an extension of the widely-used OMNeT++ simulation platform and its INET framework \cite{b5}, \cite{b6}, \cite{b7}. By leveraging OMNeT++’s modular architecture, Simu5G provides researchers with a flexible environment to model radio access, backhaul, and mobile edge computing scenarios in 5G New Radio (NR) deployments.

\subsection{Simu5G Protocol Stack Structure}
Simu5G implements a complete user plane protocol stack for 5G NR UEs and gNBs. The stack includes:
\begin{itemize}
    \item PDCP Layer: Responsible for header compression, sequence numbering, and reordering.
    \item RLC Layer: Provides ARQ-based reliable transmission (AM), unacknowledged transmission (UM), or transparent mode (TM).
    \item MAC Layer: Handles logical channel multiplexing, HARQ, and scheduling interactions.
    \item PHY Layer: Models simplified physical-layer transmission and channel models, including interference calculation and SINR-based link adaptation.
\end{itemize}

Each protocol sublayer is implemented as a dedicated OMNeT++ module, which allows users to flexibly compose network nodes (UEs, gNBs, routers, applications, etc.) using compound modules and customizable configurations.

Simu5G inherits many of its lower-layer LTE mechanisms from earlier versions of SimuLTE, with significant enhancements to model NR-specific features at the PHY and MAC layers, including numerology, multiple carriers, and beamforming support \cite{b5}.

\subsection{Missing SDAP Functionality}  
Despite its comprehensive support for most 5G NR layers, Simu5G does not currently include the Service Data Adaptation Protocol (SDAP) layer, which was newly introduced in 5G NR as specified in 3GPP TS 38.331 \cite{b1} and TS 38.413 \cite{b2}. In Simu5G’s default architecture:
\begin{itemize}
    \item Packets generated by applications are passed directly to the PDCP layer after IP processing.
    \item No SDAP header insertion or removal occurs.
    \item There is no representation of QoS Flows or QFI-based flow separation.
    \item PDCP instances are typically bound to a single default DRB without support for multiple bearers per UE.
\end{itemize}

Consequently, all QoS flows are effectively multiplexed at the application or IP level, and the simulator cannot capture the QFI-based bearer multiplexing behavior that is central to 5G NR’s QoS architecture \cite{b3}, \cite{b4}, \cite{b8}. This limits Simu5G's ability to study advanced topics such as:
\begin{itemize}
    \item Flow isolation between different QoS Flows;
    \item Per-flow latency and delay performance under mixed service classes;
    \item Bearer reconfiguration and DRB multiplexing;
    \item Interaction between SDAP and higher-layer control mechanisms such as NGAP \cite{b2}.
\end{itemize}

\subsection{Motivating the SDAP Extension}
The absence of SDAP support creates a structural modeling gap for researchers interested in studying:
\begin{itemize}
    \item Fine-grained QoS handling at the flow level;
    \item Realistic simulation of multi-QFI scenarios;
    \item The impact of DRB multiplexing on latency, throughput, and isolation;
    \item Scheduling policies that take QFI information into account at the MAC layer.
\end{itemize}

To address this gap, we propose a modular SDAP extension that integrates seamlessly with Simu5G’s existing UE and gNB protocol stacks. Our implementation introduces SDAP header processing, QFI tagging, and logical DRB mapping while remaining backward-compatible with the existing LTE-inspired bearer management framework.

\section{SDAP Extension Design} \label{section3}
In this section, we present the design and integration of the SDAP sublayer into the Simu5G simulation framework. Our extension introduces SDAP functionality into both the UE and gNB protocol stacks by inserting dedicated SDAP entities between the IP and PDCP layers. This allows proper insertion and removal of SDAP headers, per-flow QFI assignment, and logical DRB mapping, all while preserving compatibility with Simu5G’s existing LTE-inspired architecture.

To ensure maintainability and extensibility, the SDAP functionality was implemented as two dedicated modules:

\begin{itemize}

\item \textbf{NrTxSdapEntity:} responsible for SDAP header insertion and QFI multiplexing in the transmission path.

\item \textbf{NrRxSdapEntity:} responsible for SDAP header removal and QFI extraction in the reception path.
\end{itemize}
Our design leverages OMNeT++’s modular composition and INET’s flexible packet tagging and chunk-based serialization framework to achieve protocol-compliant header processing. Furthermore, a simple yet powerful QFI-to-DRB mapping mechanism is introduced to enable flow-based bearer differentiation without requiring intrusive changes to Simu5G’s lower-layer internals. The following subsections detail the architecture and functionality of the transmit and receive SDAP entities, the QFI assignment mechanism, and the logical DRB mapping approach used in our implementation.

\begin{figure}
    \centering
    \includegraphics[width=\linewidth]{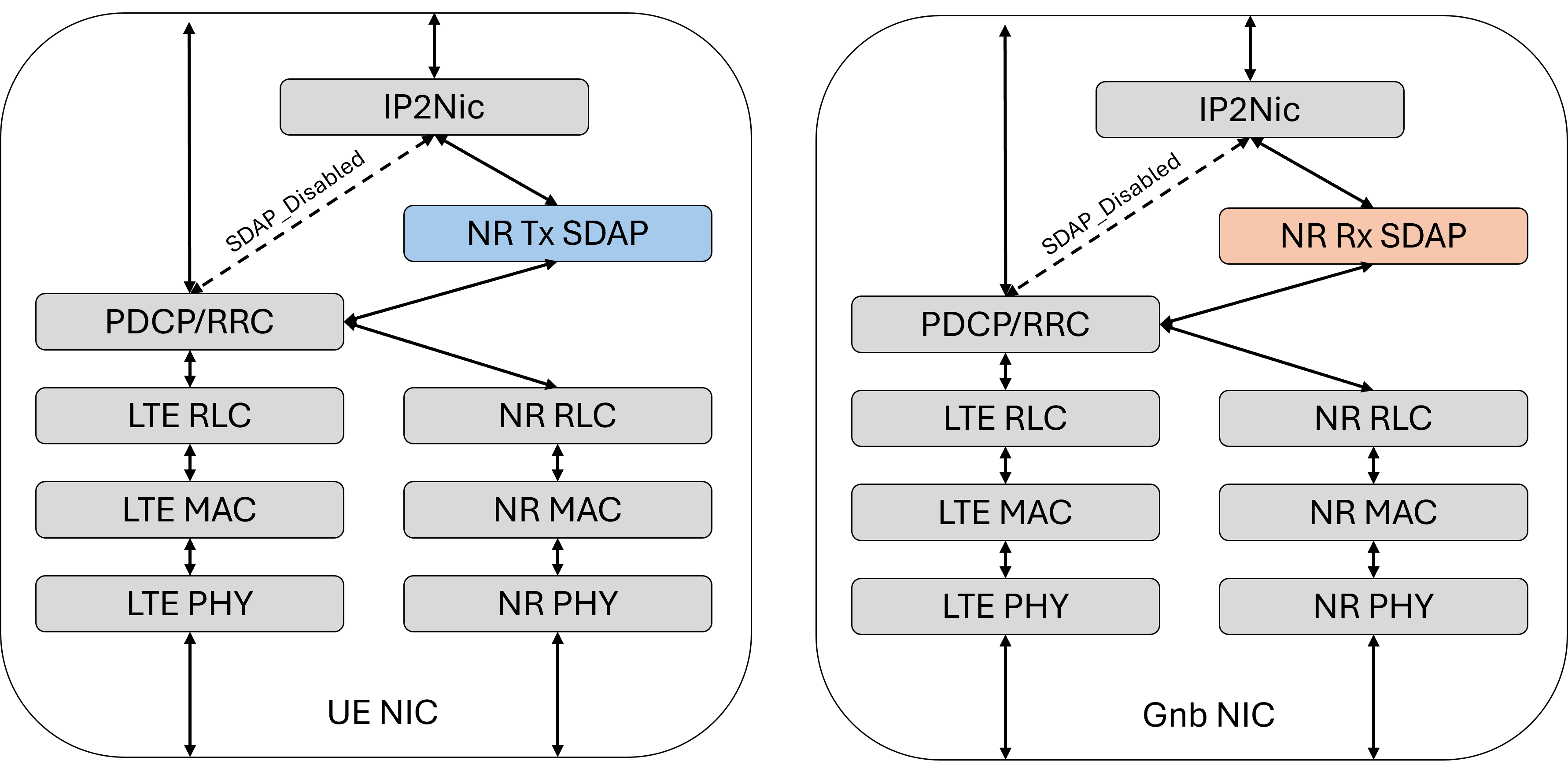}
    \caption{Integration of SDAP entities into the Simu5G protocol stack. The transmit-side (NrTxSdapEntity) and receive-side (NrRxSdapEntity) modules enable QFI-based header processing and logical DRB mapping between the IP and PDCP layers.}
    \label{fig:1}
\end{figure}

\subsection{Transmit-Side Architecture (NrTxSdapEntity)}
The transmit-side SDAP functionality is implemented within a dedicated OMNeT++ module named NrTxSdapEntity. This module sits between the IP layer and PDCP layer inside the UE and gNB protocol stacks, as shown in Fig. \ref{fig:1}. Its primary function is to process outgoing IP packets, extract the corresponding QFI, insert the SDAP header, and forward the resulting Service Data Unit (SDU) toward the PDCP layer for further processing.

The SDAP header insertion process is performed in several well-defined steps:

\textbf{1- QFI Extraction from Application Layer Tags:} To assign QFIs to different application flows, we leverage OMNeT++'s packet tagging mechanism provided by the INET framework. Each application attaches a custom QosTagReq tag to its packets, which carries the desired QFI value as a parameter. This allows different applications to dynamically assign different QFIs through simple .ini configuration, without hardcoding SDAP logic into the applications themselves.

When a packet arrives at the NrTxSdapEntity module via its DataPort\$i gate, the SDAP entity reads the attached QosTagReq tag (if present) and extracts the QFI value. If no tag is found, a default QFI of zero is assumed.

\textbf{2- Header Extraction for Correct SDAP Placement:} To maintain strict protocol compliance, the SDAP header is positioned between the transport and network layers, as specified in 3GPP TS 38.331 \cite{b1}. Therefore, before inserting the SDAP header, the module parses and removes: The outer IPv4 header and the transport header, supporting both UDP and TCP, using conditional parsing based on the protocol ID. This ensures that the SDAP header is inserted at the correct protocol layer boundary.

\textbf{3- SDAP Header Insertion:} A new NrSdapPdu chunk is created, representing the SDAP header. The extracted QFI is assigned to the QFI field of the SDAP header. The D/C bit is set to "data" mode, and the RQI is set to its default value of false. The header is then inserted at the front of the packet.

\textbf{4- Reinsertion of Transport and Network Headers:} Following SDAP header insertion, the removed transport and network headers are re-attached in reverse order. For UDP packets, both the transport header length (chunkLength) and total length field (totalLengthField) are updated to account for the additional SDAP header. The same adjustments apply to the IP header to reflect the updated total packet size.

\textbf{5- Logical DRB Mapping (Non-intrusive Multiplexing):}After inserting the SDAP header, the module consults an internal QFI-to-DRB mapping table to determine the logical DRB associated with the extracted QFI. The mapping table is populated at runtime by parsing a configuration string provided via the OMNeT++ .ini file.

Although physical DRB multiplexing is not yet implemented, this logical mapping allows the system to log DRB selections for each flow, enabling clear verification of multiplexing decisions. If a QFI is not found in the mapping table, a default DRB value of zero is used.

\textbf{6- Forwarding to PDCP:} Finally, the fully processed packet—with its SDAP header inserted—is forwarded to the PDCP layer via the stackSdap\$o gate. The PDCP layer, which operates unchanged from the original Simu5G implementation, remains fully transparent to the SDAP header due to its placement after the transport and IP headers.

\subsection{Receive-Side Architecture (NrRxSdapEntity)}
The receive-side SDAP functionality is implemented via the corresponding NrRxSdapEntity module. This module sits between the PDCP and IP layers in the reception path and is responsible for parsing incoming SDAP headers, extracting the QFI, and removing the SDAP header prior to forwarding the packet to the IP layer.

The SDAP header removal process is symmetrical to the transmit-side logic and follows these ordered steps:

\textbf{1- Header Extraction Preparation:} When a packet arrives at the NrRxSdapEntity via its stackSdap\$i gate (i.e., from PDCP), the outer IP header is first extracted using the removeAtFront<Ipv4Header>() method. Similar to the transmit path, the transport protocol ID is read from the IP header to determine whether the transport layer contains UDP or TCP headers.

\textbf{2- Transport Header Parsing:} Depending on the identified transport protocol an extraction method is invoked. This clean extraction of both IP and transport headers ensures that the SDAP header resides at the front of the remaining packet payload.

\textbf{3-  SDAP Header Parsing:}  The SDAP header then is extracted. This operation yields the embedded QFI value associated with the packet. The QFI extraction serves two key purposes:
\begin{itemize}
    \item To enable downstream modules to be aware of the QoS Flow that carried the packet.
    \item To allow DRB selection verification via the same logical mapping used at the transmit side.
\end{itemize}

\textbf{4- Logical DRB Lookup and Logging:} Similar to the transmit-side entity, NrRxSdapEntity maintains a QFI-to-DRB mapping table, which is populated using the same .ini configuration string at simulation initialization. Upon QFI extraction, the module performs a lookup in this mapping table to retrieve the corresponding DRB. This mapping is used for logging and validation purposes, ensuring symmetry with the transmit-side logic. If the extracted QFI is not found in the mapping table, a default DRB value of zero is assumed.

The logging of QFI and DRB assignments at both transmission and reception paths allows for end-to-end verification of the flow mapping behavior without requiring intrusive changes to the underlying PDCP or RLC layers.

\textbf{5- Header Reassembly and Forwarding: } Following SDAP header removal, the transport and IP headers are reinserted at the front of the packet. The totalLengthField fields of both headers are updated to reflect the removal of the SDAP header, ensuring packet consistency.

Finally, the fully reconstructed packet is forwarded to the IP layer via the DataPort\$o gate, allowing standard INET or Simu5G applications to receive the packet without any awareness of SDAP header processing.

\subsection{QFI Tagging Mechanism}
In 5G NR, the assignment of QoS Flows and their associated QFIs is typically handled by the control-plane, with QFIs communicated to the user plane via standardized signaling procedures such as NAS and RRC \cite{b1}, \cite{b2}, \cite{b3}. Since Simu5G currently does not implement full control-plane procedures for dynamic bearer and QoS Flow management, we adopt a lightweight and fully flexible approach for QFI assignment that leverages OMNeT++'s powerful packet tagging mechanism provided by the INET framework \cite{b6}, \cite{b7}.

\textbf{1- Use of Packet Tags: } OMNeT++’s INET framework allows arbitrary metadata to be attached to packets via tags, which are independent of the packet payload and protocol headers. These tags do not alter the byte stream of the packet, but rather serve as annotations that can be read and modified by various layers within the protocol stack.

\textbf{2- Custom Tag Definition: QosTagReq: } To represent the QFI information, we introduce a custom tag class named QosTagReq. This tag contains a single field representing the assigned QFI value for the packet. The QosTagReq tag is defined via OMNeT++ message definitions (QosTag.msg), extending INET's standard tagging infrastructure while remaining fully compatible with Simu5G's modular architecture.

\textbf{3- Application-Level QFI Assignment: } At the application layer (e.g., within UDP or TCP-based traffic generators), each outgoing packet is assigned a QFI value by inserting the QosTagReq tag prior to transmission. The QFI value is configured via the OMNeT++ .ini configuration file on a per-application basis. The applications are instrumented to read this configuration parameter and attach the corresponding QosTagReq tag. This mechanism allows fully independent QFI assignment for each traffic flow without requiring complex bearer configuration procedures.

\textbf{4- QFI Extraction at SDAP Layer: } Upon arrival at the NrTxSdapEntity, the SDAP module extracts the QFI value directly from the attached QosTagReq tag. If no QFI tag is present, a default QFI value of zero is assumed. This design allows full backwards compatibility for applications that are not explicitly configured with QFIs.

The proposed tagging mechanism offers several advantages:
\begin{itemize}
    \item Clean separation of application-layer and transport-layer logic.
    \item No modification to application message structures or payload formats.
    \item Fully dynamic and flexible per-flow QFI configuration via .ini file.
    \item Minimal impact on simulation scalability or performance.
    \item Seamless extensibility for more advanced SDAP scenarios.
\end{itemize}

By adopting the tag-based approach, we avoid unnecessary protocol stack coupling and enable easy experimentation with multi-QFI traffic scenarios within Simu5G.

\subsection{QFI-to-DRB Mapping Logic}

In 5G NR systems, QoS Flows identified by their respective QFIs may be multiplexed onto one or more DRBs, depending on operator policies, subscription profiles, and control-plane decisions \cite{b1}, \cite{b3}, \cite{b10}. While 3GPP specifications support both one-to-one and many-to-one QFI-to-DRB mappings, Simu5G’s current implementation does not support multiple DRBs or bearer-level multiplexing natively.

To address this gap while maintaining full compatibility with Simu5G’s existing architecture, we introduce a lightweight logical QFI-to-DRB mapping mechanism into both NrTxSdapEntity and NrRxSdapEntity. This allows us to simulate the multiplexing decision-making process of SDAP while leaving PDCP, RLC, and MAC layers unchanged.

The mapping between QFIs and DRBs is configured statically at simulation initialization through OMNeT++’s .ini file. We define a custom parameter named qfiToDrbMapping in both transmit and receive SDAP entities. The parameter supports flexible mappings and can be easily extended for additional flows.

During the initialize() function of both SDAP entities, the configuration string is parsed into an internal mapping table. The parsing logic allows clean separation of configuration handling from runtime packet processing. Each QFI-DRB pair extracted from the configuration string is inserted into the map for efficient lookup during simulation.

When the SDAP TX entity processes an outgoing packet, after extracting the QFI value from the QosTagReq tag, it performs a lookup into the mapping table to determine the associated DRB. Although actual physical DRB multiplexing is not implemented at this stage, this logical mapping allows: Logging of DRB selection decisions; Verification of correct QFI-to-DRB assignments; Future extensibility toward full DRB multiplexing. If no mapping exists for a given QFI, a default DRB of 0 is assumed.

At the receiving side, the same qfiToDrbMapping parameter is used to populate the receive-side mapping table. After SDAP header removal and QFI extraction, the DRB mapping is re-evaluated for validation and logging purposes. This symmetry ensures that DRB selection decisions remain fully traceable end-to-end, even in the absence of physical DRB separation. 

This logical QFI-to-DRB mapping mechanism is designed to: Require no modification to Simu5G's PDCP, RLC, or MAC layers. Allow researchers to experiment with multi-QFI configurations and monitor DRB allocation decisions. Serve as a foundation for future enhancements, where full per-DRB PDCP/RLC instances may be introduced.

By decoupling QFI processing from physical bearer multiplexing at this stage, we enable controlled incremental development while validating SDAP correctness.

\begin{table*}[t]
\centering
\caption{SDAP Functional Validation Checklist}
\resizebox{\linewidth}{!}{
\begin{tabular}{|l|l|l|l|}
\hline
\textbf{Validation Checkpoint} & \textbf{Module} & \textbf{Expected Behavior} & \textbf{Result} \\
\hline
QFI extraction from \texttt{QosTagReq} tag & \texttt{NrTxSdapEntity} & QFI parsed from packet tag & \checkmark Passed \\
\hline
SDAP header insertion (QFI, D/C, RQI fields) & \texttt{NrTxSdapEntity} & Header inserted between transport and IP & \checkmark Passed \\
\hline
QFI-to-DRB logical mapping & \texttt{NrTxSdapEntity} & QFI mapped to DRB via \texttt{.ini} config & \checkmark Passed \\
\hline
SDAP header removal & \texttt{NrRxSdapEntity} & Header stripped before IP forwarding & \checkmark Passed \\
\hline
QFI extraction from SDAP header & \texttt{NrRxSdapEntity} & QFI parsed from SDAP field & \checkmark Passed \\
\hline
DRB mapping verification & \texttt{NrRxSdapEntity} & DRB mapping matches TX side & \checkmark Passed \\
\hline
End-to-end packet integrity & Both & No loss, corruption, or mismatch & \checkmark Passed \\
\hline
\end{tabular}
}
\label{tab:sdap_validation}
\end{table*}

\section{Implementation Details} \label{section4}
Due to space constraints, we omit the low-level implementation details of the developed SDAP modules. The complete source code, including all modified and newly introduced components such as:
\begin{itemize}
    \item NrTxSdapEntity (transmit-side SDAP entity),
    \item NrRxSdapEntity (receive-side SDAP entity),
    \item NrSdapPdu message definitions,
    \item QosTag message definitions,
    \item Modified NED files for UE and gNB protocol stacks,
\end{itemize}
is publicly available in our GitHub repository at \cite{b9}. This implementation is fully integrated into Simu5G while preserving compatibility with existing modules. The repository contains all necessary configuration files, scenario examples, and documentation to reproduce the presented results and extend the system for further research.

\section{Evaluation} \label{eval}
We evaluated the proposed SDAP extension in terms of both functional correctness and QoS differentiation capability using the Simu5G framework.

\subsection{Simulation Setup}
To evaluate the proposed SDAP extension, we used Simu5G to simulate a standalone 5G NR deployment in a single-cell scenario. The network topology consists of one gNB located at coordinates (450 m, 300 m) and one stationary UE placed at (450 m, 350 m). The physical layer was configured with 50 resource blocks, while transmission powers were set to 40 dBm for the gNB and 26 dBm for the UE. Mobility and fading were disabled to ensure stable channel conditions and isolate protocol-level behavior.

The simulation duration was set to 20 seconds. Traffic was generated using multiple UDP-based constant bit rate (CBR) applications to model delay-sensitive services such as VoIP. Each application was configured with a unique source-destination port pair and operated at a fixed packet rate of 50 packets per second with a packet size of 160 bytes, representing a typical G.711 codec stream. Both uplink and downlink directions were evaluated separately.

Each application flow was mapped to a distinct QoS Flow Identifier (QFI) using the QosTagReq packet tagging mechanism. This was achieved by specifying a qfi parameter in the .ini configuration file for each application. The SDAP modules in the UE and gNB parsed these tags at runtime to insert or extract the appropriate QFI values.

The QFI-to-DRB mapping was defined statically in both transmit and receive SDAP entities using the following .ini string:
\begin{itemize}
    \item QFI 1 → DRB 0,
    \item QFI 5 → DRB 1,
    \item QFI 9 → DRB 2,
    \item QFI 63 → DRB 3. 
\end{itemize}

\subsection{Functional Validation}
To verify the correctness of the SDAP extension, we conducted a series of packet-level tracing and log analysis experiments. These tests focused on validating the proper operation of SDAP header insertion/removal, QFI tagging, and logical DRB mapping at both the UE and gNB sides.

During simulation, detailed debug logs were collected from both NrTxSdapEntity (transmit-side SDAP) and NrRxSdapEntity (receive-side SDAP). The logs confirmed the following functional behaviors:

\noindent \textbf{1. QFI Extraction:} Each outgoing packet was correctly assigned a QFI based on the QosTagReq attached by the application layer. For example:

\begin{verbatim} 
[TX] QFI = 5
     extracted from QosTagReq; 
\end{verbatim}

\noindent \textbf{2. SDAP Header Insertion:} The NrTxSdapEntity module inserted a compliant SDAP header, setting the QFI field, the Data/Control (D/C) bit, and disabling the RQI. The packet size and transport/IP headers were adjusted accordingly:

\begin{verbatim}
[TX] Inserted SDAP header
     with QFI = 5; 
\end{verbatim}

\noindent \textbf{3.Logical DRB Mapping:} The QFI was mapped to a DRB using the configured table. This mapping was logged for each packet:

\begin{verbatim}
[TX] Selected DRB = 1 for QFI = 5.
\end{verbatim}

\noindent \textbf{4.Symmetric Reception:} The NrRxSdapEntity successfully removed the SDAP header and extracted the embedded QFI. DRB mapping was re-evaluated to confirm consistency with the transmit side:

\begin{verbatim}
[RX] Extracted QFI = 5
[RX] Mapped DRB = 1
\end{verbatim}

This end-to-end validation was performed for all configured QFIs (1, 5, 9, and 63), in both uplink and downlink directions. Across all flows, the SDAP modules correctly handled header insertion and removal, QFI tagging, and DRB mapping without introducing packet loss, flow misclassification, or DRB mismatches. Packet integrity was preserved throughout the transmission and reception paths.

As summarized in Table \ref{tab:sdap_validation}, the results confirm that the proposed SDAP extension adheres to 3GPP specifications and integrates seamlessly into Simu5G. Crucially, this functionality is achieved without requiring any modifications to the PDCP, RLC, or MAC layers.

\begin{table}[t]
\centering
\caption{Per-Flow Latency With and Without SDAP}
\footnotesize
\begin{tabular}{|c|c|c|c|}
\hline
\textbf{Scenario} & \textbf{QFI} & \textbf{Mean Latency (ms)} & \textbf{Std Dev (ms)} \\
\hline
Without SDAP & — & 18.3 & 4.5 \\ \hline
With SDAP & 1 & 11.9 & 1.4 \\
With SDAP & 5 & 13.2 & 1.7 \\
With SDAP & 9 & 15.0 & 2.0 \\
With SDAP & 63 & 12.1 & 1.5 \\
\hline
\end{tabular}
\label{tab:qfi_latency}
\end{table}

\subsection{QoS Flow Differentiation Results}
To demonstrate the impact of the proposed SDAP extension on flow-level QoS differentiation, we compared per-flow latency measurements with and without SDAP enabled. In the baseline configuration, Simu5G does not implement SDAP and multiplexes all traffic flows over a single default bearer. In contrast, our SDAP-enhanced setup uses distinct QFIs mapped to logical DRBs to isolate flows at the SDAP layer.

Four UDP applications were configured to generate constant bit rate traffic, each tagged with a unique QFI (1, 5, 9, or 63). In both scenarios, packet-level latency was recorded using Simu5G’s tracing infrastructure. Table~\ref{tab:qfi_latency} presents the average latency and standard deviation observed for each flow.

The results indicate that SDAP-based QFI separation reduces overall latency and significantly lowers jitter (as reflected by the standard deviation). This is attributed to the logical flow isolation enabled by SDAP, which prevents cross-flow interference at the application and transport layers. 

These findings confirm that the proposed extension enables Simu5G to simulate per-QFI flow separation effects, which are critical for studying scheduling policies, delay-sensitive applications, and heterogeneous 5G services in future research.

\section{Discussion} \label{section5}
The proposed SDAP extension equips Simu5G with critical functionality for 3GPP-compliant QoS Flow management. By introducing SDAP header processing, QFI-based flow tagging, and configurable logical DRB mapping, our implementation enables realistic simulation of per-flow differentiation—capabilities that were previously unavailable in the Simu5G framework.

The modular design ensures seamless integration with existing Simu5G components without modifying lower-layer modules such as PDCP, RLC, or MAC. This design decision preserves compatibility and minimizes the learning curve for researchers and developers extending the platform.

Through functional validation and latency-based differentiation experiments, we demonstrate that the SDAP layer correctly enables flow isolation and reduces cross-flow interference. As confirmed by our evaluation, QFI-separated flows exhibit lower latency variance compared to Simu5G’s default bearer model, validating the practical value of this enhancement.

From a research perspective, this extension opens up a wide range of new simulation scenarios. These include:
\begin{itemize}
    \item Evaluation of per-QFI scheduling algorithms at the MAC layer,
    \item Analysis of heterogeneous traffic classes such as URLLC, eMBB, and mMTC,
    \item Investigation of QoS-aware routing and resource allocation strategies,
    \item Simulation of industrial and real-time applications requiring strict latency bounds.
\end{itemize}

It is important to emphasize that the present implementation performs logical multiplexing only. The QFI-to-DRB mapping implemented within SDAP allows us to track bearer allocation decisions without requiring intrusive modifications to the PDCP, RLC, or MAC layers in Simu5G. This non-intrusive approach makes the solution immediately deployable on top of existing Simu5G releases.

Nevertheless, certain 3GPP features are intentionally left for future work. In particular:
\begin{itemize}
    \item Physical DRB multiplexing is not yet supported. All QoS flows are still forwarded through a single PDCP/RLC stack instance in the current architecture.
    \item Control-plane signaling for dynamic QFI allocation and bearer management is not implemented, as this would require full NAS and NGAP modeling.
    \item Support for advanced SDAP functionalities such as RQI, control PDU formats, and DRB failover mechanisms is not included at this stage.
\end{itemize}

Despite these limitations, our design lays the groundwork for safe and incremental extension of Simu5G toward full multi-DRB support. The clean modular separation between SDAP and lower layers ensures that physical DRB multiplexing can be added in the future without requiring any major redesign of the SDAP entities described here.

\section{Future Work and Conclusion} \label{section6}
In this paper, we presented a modular and standards-compliant implementation of the Service Data Adaptation Protocol (SDAP) for Simu5G, introducing support for SDAP header processing, QFI-based flow differentiation, and logical DRB mapping. This enhancement enables, for the first time, simulation of multiple QoS flows per UE in accordance with 3GPP 5G NR specifications—without requiring changes to existing lower-layer modules.

Our extension integrates seamlessly with Simu5G's LTE-inspired architecture and offers a lightweight yet flexible mechanism for configuring QFI and DRB behavior via .ini files. Through comprehensive validation and performance evaluation, we demonstrated that the system correctly implements per-QFI flow separation and reduces latency variability compared to Simu5G’s default bearer model.

As future work, we plan to extend the current design to support:
\begin{itemize}
    \item Physical DRB multiplexing, including per-DRB PDCP and RLC instances,
    \item Dynamic control-plane signaling for runtime QFI assignment and DRB reconfiguration,
    \item Advanced SDAP features, such as Reflective QoS Indicator (RQI) handling and control PDU formats,
    \item Integration with QoS-aware MAC schedulers to evaluate cross-layer policies.
\end{itemize}

All source code, configuration files, and evaluation artifacts are made publicly available to encourage adoption and further research within the 5G simulation community. By bridging this critical gap in Simu5G, our SDAP extension lays the groundwork for more realistic and fine-grained 5G NR simulations—enabling new research directions across QoS, scheduling, and industrial network modeling.

\balance
\end{document}